\documentclass[preprint]{revtex4}
\usepackage{amssymb}
\begin{document}

\title{Generalization of Shannon-Khinchin Axioms \\
to Nonextensive Systems and the Uniqueness Theorem}

\author{Hiroki Suyari} 

\affiliation{%
Department of Information and Image Sciences,
Faculty of Engineering, Chiba University\\
1-33, Yayoi-cho, Inage-ku, Chiba-shi, Chiba, 263-8522 Japan
}%

\email{suyari@tj.chiba-u.ac.jp}
\date{\today\\ \bigskip\bigskip\bigskip\bigskip\bigskip}

\begin{abstract}
The Shannon-Khinchin axioms are generalized to nonextensive systems
and the uniqueness theorem for the nonextensive entropy is proved
rigorously.
In the present axioms, Shannon additivity is used as additivity
in contrast to pseudoadditivity in Abe's axioms.
The results reveal that Tsallis entropy is the simplest among
all nonextensive entropies which can be obtained from the generalized
Shannon-Khinchin axioms. 
\end{abstract}
\maketitle
\section{Introduction}

Tsallis entropy, which was first introduced by Tsallis in 1988 \cite{Ts88},
is defined by
\begin{equation}
S_q^T\left( {p_1,\ldots ,p_n} \right) 
\equiv{{1-\sum\limits_{i=1}^n{p_i^q}} \over {q-1}},
\label{Tsallisentropy}
\end{equation}
for a given probability distribution $\{p_i\}_{i=1...n}$, where $n$ is a
number of accessible configurations, $q$ is a real parameter assumed
to be positive, and Boltzmann's constant $k$ has been set equal to
unity.
     
Tsallis entropy is a one-parameter generalization of Shannon entropy
in the sense that
\begin{equation}
\mathop {\lim }\limits_{q\to 1}S_q^T=S_1\mathop 
\equiv -\sum\limits_{i=1}^n {p_i\ln p_i}.
\label{Shannonentropy}
\end{equation}
The characteristic property of Tsallis entropy is called 
{\it pseudoadditivity}, i.e. 
\begin{equation}
S_q^T\left( A,B \right)=S_q^T\left( A \right)+S_q^T\left( B \right)+
\left( 1-q \right)S_q^T\left( A \right)S_q^T\left( B \right)
\label{pseudoadditivity}
\end{equation}
holds true for two mutually independent finite event systems $A$ and $B$ 
\begin{equation}
A=\left( {\matrix{{A_1}&\ldots &{A_n}\cr
{p_1^A}&\ldots &{p_n^A}\cr
}} \right),\quad B=\left( {\matrix{{B_1}&\ldots &{B_m}\cr
{p_1^B}&\ldots &{p_m^B}\cr
}} \right),
\label{ABsystem}
\end{equation}
\begin{equation}
\hbox{such that}\quad p_{ij}^{AB}=p_i^Ap_j^B\quad  {\hbox{for any}}\;\;{i=1,\;\ldots ,\;n}
\hbox{ and } {j=1,\;\ldots ,\;m}.
\label{factor}
\end{equation}
The property (\ref{pseudoadditivity}) represents nonextensive
additivity of Tsallis entropy, i.e. Tsallis entropy is a {\it
nonextensive entropy}. When $q=1$ the pseudoadditivity
(\ref{pseudoadditivity}) reverts to the standard additivity, so that
Shannon entropy is considered to be {\it extensive entropy} in
contrast to Tsallis entropy. 

Introduction of Tsallis entropy
opened a new research area in statistical physics, providing an
advantageous generalization of traditional Boltzmann-Gibbs statistical
mechanics \cite{AO01}. The generalization enables us to
find a consistent treatment of dynamics in many nonextensive physical systems
such as long-range interactions,
long-time memories, and multi-fractal structures,
which cannot be coherently explained
within the conventional Boltzmann-Gibbs statistics \cite{AO01}.
Thus Tsallis entropy inspires many physicists to establish a
generalized Boltzmann-Gibbs statistical mechanics leading to numerous
applications \cite{AO01,CT91}. 

Along with the rapid progress of research in the generalized
Boltzmann-Gibbs statistical mechanics, some axioms of Tsallis entropy
(\ref{Tsallisentropy}) were presented as generalization of
Shannon-Khinchin axioms \cite{Sa97,Ab00}. We proceed with a discussion
of the Shannon-Khinchin axioms and their possible generalized counterparts.

Let $\Delta _n$ be defined by the $n$-dimensional simplex:
\begin{equation}
\Delta _n\mathop\equiv \left\{ {\left( {p_1,\ldots ,p_n} \right)
\left|\;{p_i\ge 0,\;\sum\limits_{i=1}^n {p_i}=1} \right.} \right\}.
\label{simplex}
\end{equation}
The Shannon-Khinchin axioms \cite{Kh57} are given by the
following four conditions:
\begin{description}
\item{[SK1]} {\it continuity}: for any $n\in \mathbb{N}$ the function $S_1\left(
p\right)$  is continuous with respect to $p\in \Delta _n$,
\item{[SK2]} {\it maximality}: for given $n\in \mathbb{N}$ and 
for $\left( {p_1,\;\ldots ,\;p_n} \right)\in \Delta_n$,
the function $S_1\left( {p_1,\;\ldots ,\;p_n} \right)$
takes its largest value for 
$p_i={1 \over n}\;\left( {i=1,\;\ldots ,\;n} \right)$,
\item{[SK3]} {\it additivity (Shannon additivity)}: 
if %
\begin{equation}
p_{ij}\ge 0,\; p_i=\sum\limits_{j=1}^{m_i} {p_{ij}}\;\;
\hbox{for any} \;\; i=1,\ldots,n\;\hbox{and}\;j=1,\ldots,m_i,\hbox{and}\
\sum\limits_{i=1}^n {p_i}=1,
\label{condition0}
\end{equation}
then the following equality holds:
\begin{equation}
S_1\left( {p_{11},\ldots,p_{nm_n}} \right)
=S_1\left( {p_1,\ldots,p_n}\right)
+\sum\limits_{i=1}^n {p_iS_1\!\left( {{{p_{i1}}
\over {p_i}},\ldots,{{p_{im_i}}\over {p_i}}} \right)},
\label{Shannonadditivity}
\end{equation}
\item{[SK4]} {\it expandability}:
$S_1\left( {p_1,\ldots,p_n,0} \right)=S_1\left( {p_1,\ldots,p_n} \right)$.
\end{description}
In order to discriminate between the pseudoadditivity (\ref{pseudoadditivity})
and the above additivity property (\ref{Shannonadditivity}),
we call the additivity (\ref{Shannonadditivity}) {\it Shannon additivity}
throughout the paper. 

Until now,
two sets of axioms for Tsallis entropy
were presented as generalizations of the above Shannon-Khinchin axioms
\cite{Sa97,Ab00}. In the former axioms presented in \cite{Sa97}, the four conditions are given
as axioms for Tsallis entropy $S_q^T\left( p\right)$:
[Sa1] continuity - same as [SK1] for $S_q^T$, 
[Sa2] increasing monotonicity - when $p_i={1 \over n}\;\left( {i=1,\;\ldots ,\;n} \right)$,
$S_q^T$ is a monotonic increasing function of $n$ for any $q\in \mathbb{R}^+$,
[Sa3] pseudoadditivity - the equality (\ref{pseudoadditivity}) holds, 
and [Sa4] generalized Shannon additivity - under the same constraints as (\ref{condition0}),
the following equality holds
\begin{equation}
S_q^T\left( {p_{11},\ldots,p_{nm_n}} \right)
=S_q^T\left( {p_1,\ldots,p_n}\right)
+\sum\limits_{i=1}^n {p_i^qS_q^T\!\left( {{{p_{i1}}
\over {p_i}},\ldots,{{p_{im_i}}\over {p_i}}} \right)}
\label{q_Tsallisadditivity}.
\end{equation}
This condition (\ref{q_Tsallisadditivity}) is
a generalization of the original one that appeared in \cite{Sa97},
which is a case  of (\ref{q_Tsallisadditivity}) when $n=2$.
This generalization is a straightforward one \cite{AO01,Ts95}.

In \cite{Sa97}, the uniqueness theorem stating that
the only function satisfying all requirements [Sa1]$\sim$[Sa4] is
Tsallis entropy is proved, but actually these four axioms are verbose.
In fact, it is sufficient to have two axioms, pseudoadditivity [Sa3] and
generalized Shannon additivity [Sa4], to determine Tsallis entropy
uniquely, as has been shown in our paper \cite{Su01b}.
Thus the four axioms [Sa1]$\sim$[Sa4] in \cite{Sa97} are clearly redundant.

On the other hand, 
in the latter axioms presented in \cite{Ab00}, the three conditions
given as axioms for Tsallis entropy $S_q^T$ are [Ab1] continuity and
maximality - same conditions as [SK1] and [SK2] for $S_q^T$,
[Ab2] pseudoadditivity using the conditional entropy, 
and [Ab3] expandability - same condition as [SK4] for $S_q^T$. 
The uniqueness theorem for Tsallis entropy is also proved.


%
%
 
In this paper, the Shannon-Khinchin axioms are generalized to
nonextensive systems and the uniqueness theorem for nonextensive entropies 
including Tsallis entropy is proved rigorously.
In our axioms, Shannon additivity is used as additivity
in contrast to pseudoadditivity using the conditional entropy
in Abe's axioms \cite{Ab00}.

It was found recently that the original Tsallis entropy
(\ref{Tsallisentropy}) needs to be renormalized to satisfy the form
invariance of the maximum entropy principle or the pseudoadditivity
\cite{RA99,Su01a}. 
The normalized Tsallis entropy is defined by 
\begin{equation}
\hat S_q^T\left( {p_1,\ldots ,p_n} \right)\mathop\equiv 
{{1-\sum\limits_{i=1}^n {p_i^q}} \over {\left( {q-1} \right)\sum\limits_{j=1}^n
{p_j^q}}}.
\label{n_Tsallisentropy}
\end{equation}
In order to avoid confusion, we call $S_q^T$ given by (\ref{Tsallisentropy}) and $\hat S_q^T$ given by
(\ref{n_Tsallisentropy}) {\it original Tsallis entropy}
and {\it normalized Tsallis entropy} respectively.

In the next section, the axioms for nonextensive entropy
including the original Tsallis entropy are presented,
and the uniqueness theorem is proved rigorously.

\section{Generalized axioms and Uniqueness Theorem for Nonextensive Entropy}

Generalized Shannon-Khinchin axioms and the uniqueness theorem
for nonextensive entropy are given below.

\bigskip
Let $\Delta _n$ be a $n$-dimensional simplex
defined by (\ref{simplex}).
The following axioms [N1]$\sim $[N4] determine
the function $S_q:\Delta _n\to \mathbb{R}^+$ such that
\begin{equation}
S_q\left( {p_1,\ldots ,p_n} \right)=
{{{1-\sum\limits_{i=1}^n {p_i^q}} }\over{\phi \left( q \right)}},
\label{theorem_Tsallis}
\end{equation}
where $\phi \left( q \right)$ satisfies properties
(i) $\sim$ (iv):
\begin{eqnarray}
&&\hspace{-1.2cm}\hbox{(i)}\;
\phi \left( q \right)\;\left\{ {\matrix{{>0,}\hfill
&{\hbox{if}\;\;q>1}\hfill
\cr {<0,}\hfill &{\hbox{if}\;\;0\le q<1}\hfill \cr
}} \right.,
\label{conditionphi1}\\
&&\hspace{-1.2cm}\hbox{(ii)}\;
\hbox{$\phi \left( q \right)$ is differentiable with
respect to 
$q\in \mathbb{R}^+$},\label{conditionphi2}\\ 
&&\hspace{-1.2cm}\hbox{(iii)}\;
\mathop {\lim }\limits_{q\to 1}
{{d\phi \left( q \right)} \over {dq}}=1,
\label{conditionphi3}\\
&&\hspace{-2.0cm}\hbox{and}\nonumber\\
&&\hspace{-1.2cm}\hbox{(iv)}\;
\mathop {\lim }\limits_{q\to 1}
\phi \left( q \right)=\phi \left( 1\right)\!=\!0,
\quad\phi \left( q \right)\ne0\; \left( {q\!\ne\!1} \right).
\label{conditionphi4}
\end{eqnarray}
\begin{description}
\item{[N1]} $S_q$ is continuous in $\Delta _n$ and $q\in \mathbb{R}^+$,
\item{[N2]} 
For any $q\in \mathbb{R}^+$ and any $n\in \mathbb{N}$,
\begin{equation}
S_q\left( p \right)\le S_q\left( {1 \over n}, \ldots, {1 \over n} \right)
\quad \hbox{for any}\;{p\in \Delta _n},
\end{equation}
\item{[N3]}
Under the constraints (\ref{condition0}),
the following equality holds:
\begin{equation}
S_q\left( {p_{11},\ldots,p_{nm_n}} \right)
=S_q\left( {p_1,\ldots,p_n}\right)
+\sum\limits_{i=1}^n {p_i^qS_q\!\left( {{{p_{i1}}
\over {p_i}},\ldots,{{p_{im_i}}\over {p_i}}} \right)},
\label{q_Shannonadditivity}
\end{equation}
\item{[N4]}
\begin{equation}
\mathop {\lim }\limits_{q\to 1}S_q=S_1
=-\sum\limits_{i=1}^n {p_i\ln p_i}.
\end{equation}
\end{description}
\bigskip

Note that the Shannon-Khinchin axioms and ours differ in the formulation of
[N3] and [N4]. [N3] is a slight generalization of the Shannon additivity
(\ref{Shannonadditivity}) in the sense that instead of
the exponent \lq\lq 1" of the coefficient $p_i$ in front of
$S_q\left( {{{p_{i1}} \over {p_i}},\ldots,{{p_{im_i}}
\over {p_i}}} \right)$, any positive real number $q$ in $p^q_i$ can
be taken. [N4] shows that Tsallis entropy is a one-parameter
generalization of the Shannon entropy.
[N4] can be replaced by the following expandability:

\medskip
[N4*] For any $q\in \mathbb{R}^+$ and any $\left( {p_1,\ldots ,p_n}
\right)\in\Delta _n$,
\begin{equation}
S_q\left({p_1,\ldots,p_n,0}\right)=S_q\left( {p_1,\ldots,p_n}\right).
\end{equation}
Furthermore, in case of $q=1$, the axioms [N1]$\sim $[N3] and [N4*] coincide with
the Shannon-Khinchin axioms \cite{Kh57,SW63}.
Therefore when $q=1$ we can obtain Shannon entropy from 
[N1]$\sim $[N3] and [N4*].
In the present axioms, for simplicity, [N4] is used instead of [N4*].

The proof is given as follows.

Consider the case of $q\ne 1$.
In the axiom [N3], we apply the following special case:
for any $i=1,\ldots ,n$ and $j=1,\ldots ,m_i$,
if we take 
\begin{equation}
m=m_1=\ldots =m_n\quad\hbox{and}\quad
p_{ij}={1 \over {mn}},
\end{equation}
then the identity (\ref{q_Shannonadditivity}) can be written as
\begin{equation}
S_q\left( {1 \over {mn}},\ldots,{1 \over {mn}} \right)
=S_q\left( {1 \over {n}},\ldots,{1 \over {n}}\right)
+\sum\limits_{i=1}^n
{{1 \over {n^q}}S_q\left( {1 \over {m}},\ldots,{1 \over {m}}\right)}.
\label{specialadditivity}
\end{equation}
Let $f_q\left( n \right)$ be defined by
\begin{equation}
f_q\left( n \right)\mathop\equiv 
S_q\left( {{1 \over n},\ldots ,{1\over n}} \right),
\label{f_q}
\end{equation}
so the equation (\ref{specialadditivity}) becomes
\begin{equation}
f_q\left( {mn} \right)=f_q\left( n \right)+n^{1-q}f_q\left( m \right).
\label{f(nm)}
\end{equation}
Exchanging the variables $m$ and $n$ in (\ref{f(nm)}), we have
\begin{equation}
f_q\left( n \right)+n^{1-q}f_q\left( m \right)
=f_q\left( m \right)+m^{1-q}f_q\left( n\right).
\end{equation}
This can be rearranged to give 
\begin{equation}
{{f_q\left( n \right)} \over {1-n^{1-q}}}={{f_q\left( m \right)} \over {1-m^{1-q}}}.
\end{equation}
This identity holds for any $m,n\in \mathbb{N}$, and it depends on the variable $q$
only. Thus, there exists a function $\phi \left( q \right)$ such that
\begin{equation}
f_q\left( n \right)={{1-n^{1-q}} \over {\phi \left( q \right)}},
\label{f(n)}
\end{equation}
where $\phi \left( q \right)$ is a function of $q\in \mathbb{R}^+$,
satisfying (\ref{conditionphi1})$\sim$(\ref{conditionphi4}).
The properties (\ref{conditionphi1})$\sim$(\ref{conditionphi4}) are
required for $\phi \left(q \right)$ to satisfy the other axioms [N2]
and [N4]. 

If all $p_i$ in $\left( {p_1,\ldots ,p_n} \right)\in \Delta _n$
are rational numbers, then for every $p_i$ there exist nonnegative
integers $m_i\in \mathbb{N}$ satisfying
\begin{equation}
p_i={{m_i} \over {\sum\limits_{i=1}^n {m_i}}}
\quad\left( {i=1,\;\ldots ,\;n} \right).
\label{p_i}
\end{equation}
Here if we take $p_{ij}$ as
\begin{equation}
p_{ij}={1 \over {\sum\limits_{i=1}^n {m_i}}}
\label{p_ij}
\end{equation}
for any $i=1,\ldots ,n$ and $j=1,\ldots ,m_i$,
then from (\ref{p_i}) we have $p_{i}={\sum\limits_{j=1}^{m_i} {p_{ij}}}$.
Applying these formulas (\ref{p_i}) and (\ref{p_ij})
to (\ref{q_Shannonadditivity}), the identity
(\ref{q_Shannonadditivity}) can be reduced to
\begin{equation}
S_q\left( {{1 \over {\sum\limits_{i=1}^n {m_i}}},\ldots ,{1 \over
{\sum\limits_{i=1}^n {m_i}}}} \right)
=S_q\left( {p_1,\ldots ,p_n}\right)
+\sum\limits_{i=1}^n {p_i^qS_q\left( {{1 \over{m_i}},\ldots ,{1 \over {m_i}}}
\right)},
\end{equation}
that is, using $f_q\left(n\right)$ defined in (\ref{f_q}),
\begin{equation}
f_q\left( {\sum\limits_{i=1}^n {m_i}} \right)
=S_q\left( {p_1,\ldots ,p_n} \right)+\sum\limits_{i=1}^n {p_i^qf_q\left( {m_i} \right)}.
\label{sumfn}
\end{equation}
Substitution of $f_q\left(n\right)$ obtained in (\ref{f(n)})
into (\ref{sumfn}) yields
\begin{eqnarray}
S_q\left( {p_1,\ldots ,p_n} \right)
&&=f_q\left( {\sum\limits_{i=1}^n {m_i}} \right)
-\sum\limits_{i=1}^n {p_i^qf_q\left({m_i} \right)}\nonumber\\
&&={{1-\sum\limits_{i=1}^n {p_i^q}} \over {\phi \left( q
\right)}}+{{\sum\limits_{i=1}^n {p_i^qm_i^{1-q}}-\left( {\sum\limits_{i=1}^n {m_i}}
\right)^{1-q}} \over {\phi \left( q \right)}}.
\label{result0}
\end{eqnarray}
The identity (\ref{p_i}) implies
\begin{equation}
\sum\limits_{i=1}^n {p_i^qm_i^{1-q}}
=\left( {\sum\limits_{i=1}^n {m_i}} \right)^{1-q},
\end{equation}
so that (\ref{result0}) becomes
\begin{equation}
S_q\left( {p_1,\ldots ,p_n} \right)
={{1-\sum\limits_{i=1}^n {p_i^q}} \over {\phi \left( q\right)}}.
\label{result1}
\end{equation}
The continuity [N1] of $S_q$ and the fact that any real number can be
approximated by rational numbers with any precision enable us to
validate the above formula for any $\left( {p_1,\ldots,p_n} \right)\in
\Delta _n$. 

Here $1-\sum\limits_{i=1}^n {p_i^q}$ in the right hand side of (\ref{result1})
is clearly a concave function with respect to
$\left( {p_1,\ldots ,p_n} \right)\in\Delta_n$ when $q >1$,
and a convex function when $0\le q <1$.
Moreover, the obtained entropy (\ref{result1}) is clearly a symmetric function
with respect to its arguments $p_1,\ldots,p_n$.
Thus the axiom [N2] requires the property (\ref{conditionphi1}) 
for $S_q\left( {p_1,\ldots ,p_n} \right)$ given by (\ref{result1}),
because the property (\ref{conditionphi1}) makes
$S_q\left( {p_1,\ldots ,p_n} \right)$ given by (\ref{result1})
a concave function on $\Delta_n$ for any fixed $q\in \mathbb{R}^+$.

Substitution of (\ref{result1}) into [N4] implies 
\begin{equation}
\mathop {\lim }\limits_{q\to 1}
{{{1-\sum\limits_{i=1}^n {p_i^q}}}\over{\phi \left( q \right)}}
=-\sum\limits_{i=1}^n {p_i\ln p_i}.
\label{limit1}
\end{equation}
Clearly $\mathop {\lim }\limits_{q\to 1}\left( {1-\sum\limits_{i=1}^n {p_i^q}}
\right)=0$, so that the axiom [N4] requires the property (\ref{conditionphi4})
for $\phi \left( q \right)$.
Moreover, the numerator ${1-\sum\limits_{i=1}^n {p_i^q}}$
is obviously differentiable with respect to $q$
and its first derivative is $-\sum\limits_{i=1}^n {p_i^q\ln p_i}$.
Thus the axiom [N4] requires the property (\ref{conditionphi2})
and (\ref{conditionphi3}) for $\phi \left( q \right)$.
Accordingly, we can apply l'H$\hat{o}$pital's rule to the left hand side of (\ref{limit1}):
\begin{equation}
\mathop {\lim }\limits_{q\to 1}{-{\sum\limits_{i=1}^n {p_i^q\ln p_i}} \over 
{{{d\phi
\left( q \right)} \over {dq}}}}=-\sum\limits_{i=1}^n {p_i\ln p_i}.
\end{equation}
Thus [N4] is satisfied as required.

Therefore, given a certain function $\phi$, we can
uniquely determine nonextensive entropy (\ref{theorem_Tsallis}). 
Thus the proof is complete.

\bigskip
Note that
there are many functions $\phi\left( q \right)$ satisfying
all properties (\ref{conditionphi1}) $\sim$ (\ref{conditionphi4}).
For instance, 
\begin{equation}
\phi \left( q \right)={{\left( {q-1} \right)\left( {q^2+1} \right)}
\over 2}
\end{equation}
is one such function, and
\begin{equation}
\phi \left( q \right)= q-1
\end{equation}
is the simplest one.
Therefore, the original Tsallis entropy (\ref{Tsallisentropy})
is the {\it simplest} of all nonextensive entropies
which can be obtained from the generalized Shannon-Khinchin axioms
[N1]$\sim$[N4].

In particular, when $\phi \left( q \right)= 1-2^{1-q}$,
$S_q$ given by (\ref{theorem_Tsallis}) is exactly the same as
Havrda-Charvat entropy $S_q^{HC}$ \cite{HC67} 
or Dar\'oczi entropy $S_q^{D}$ \cite{Da70}:
\begin{equation}
S_q^{HC}\left( {p_1,\ldots ,p_n} \right)
=S_q^{D}\left( {p_1,\ldots ,p_n} \right)\;\mathop\equiv \;
{{1-\sum\limits_{i=1}^n {p_i^q}} \over {1-2^{1-q}}}.
\label{Daroczi}
\end{equation}
Thus Havrda-Charvat entropy and Dar\'oczi entropy are also particular
examples of the nonextensive entropies.
Note that $\phi \left( q \right)=1-2^{1-q}$ does not satisfy the condition 
(\ref{conditionphi3}) because of the difference between $2$ and $e$
as base of logarithm used in each formulation.

It is easy to verify that $S_q$ obtained in our uniqueness theorem satisfies the additivity property
\begin{equation}
S_q\left( A,B \right)=S_q\left( A \right)+S_q\left( B \right)
-\phi \left( q \right)S_q\left( A \right)S_q\left( B \right)
\label{q_pseudoadditivity}.
\end{equation}
When $\phi \left( q \right)=q-1$, this equality
(\ref{q_pseudoadditivity}) takes the form of (\ref{pseudoadditivity}).
Thus the form (\ref{q_pseudoadditivity}) is a generalization of the
pseudoadditivity given by (\ref{pseudoadditivity}).
In this way the pseudoadditivity (\ref{pseudoadditivity})
or (\ref{q_pseudoadditivity}) of 
the nonextensive entropy can be derived naturally from the slight
generalization of the Shannon-Khinchin axioms.

\section{Conclusion}

There exist two main methods for axiomatic definition of Shannon entropy.
The first is to introduce inherent properties as axioms for Shannon
entropy, as done by Shannon in 1948 and Khinchin in 1953 \cite{Kh57,SW63}. 
The other way is to treat satisfactory properties as axioms for
information content of Shannon entropy, which was used in many
references such as \cite{VO79}. This method is based on the fact
that Shannon entropy is given by an expectation of information content.

In this paper we present generalized Shannon-Khinchin axioms for
nonextensive entropy and prove the uniqueness theorem rigorously
in accordance with the Shannon-Khinchin approach.
These results reveal that the Tsallis entropy is the simplest
of all nonextensive entropies
which can be obtained from the generalized Shannon-Khinchin axioms.
Moreover, the remaining problem such as redundancy
in the previously presented axioms in \cite{Sa97} is
successfully solved.

On the other hand, the latter way to define nonextensive entropy
by means of information content has been already presented in our paper \cite{Su01a}.
Thus the two manners to define nonextensive entropy
as similar to Shannon entropy are clearly observed in nonextensive
systems framework.
Comparing these two ways to define nonextensive entropy, we can point
out certain advantages of using information content over the
Shannon-Khinchin approach. For instance, the latter allows systematic
introduction of other entropies \cite{Su01a}.
In our forthcoming paper, another application of information content
will be presented for the formalism of Tsallis information theory.

\section*{Acknowledgments}
The author would like to thank Prof. Yoshinori Uesaka
and Prof. Makoto Tsukada for fruitful comments and discussions.



\medskip
\begin{thebibliography}{99}
\medskip
%
\bibitem{Ts88}C. Tsallis,
J. Stat. Phys., vol. 52, pp. 479-487, 1988.
%
\bibitem{AO01}C. Tsallis et al.,
{\it Nonextensive Statistical Mechanics and Its Applications},
edited by S. Abe and Y. Okamoto
(Springer-Verlag, Heidelberg, 2001);
see also the comprehensive list of references at 
\hbox{http://tsallis.cat.cbpf.br/biblio.htm}
%
\bibitem{CT91}E.M.F. Curado and C. Tsallis,
J. Phys. A, vol. 24, pp. L69-L72, 1991;
Corrigenda, vol. 24, pp. 3187, 1991;
vol. 25, pp. 1019, 1992.
%
\bibitem{Sa97}R.J.V. dos Santos,
J. Math. Phys., vol. 38, pp. 4104-4107, 1997.
%
\bibitem{Ab00}S. Abe,
Phys. Lett. A, vol. 271, pp. 74-79, 2000.
%
\bibitem{Kh57}A.I. Khinchin,
{\it Mathematical Foundations of Information Theory},
Dover, New York, 1957.
%
\bibitem{SW63}C.E. Shannon and W. Weaver,
{\it The Mathematical Theory of Communication},
University of Illinois Press, Urbana, 1963.
%
\bibitem{Ts95}C. Tsallis,
Chaos, Solitons and Fractals, vol. 6, pp. 539-559, 1995.
%
\bibitem{Su01b}H. Suyari,
Three classes of nonextensive entropies
characterized by Shannon additivity and pseudoadditivity,
(math-ph/0205001)
%
\bibitem{RA99}A.K. Rajagopal and S. Abe,
Phys. Rev. Lett., vol. 83, pp. 1711-1714, 1999.
%
\bibitem{Su01a}H. Suyari,
Nonextensive entropies derived from form invariance of pseudoadditivity,
to appear in Phys. Rev. E (cond-mat/0108274)
%
\bibitem{VO79}A.J. Viterbi and J.K. Omura,
{\it Principles of digital communication and coding}, 
(McGraw-Hill, 1979);
T.M. Cover and J.A. Thomas,
{\it Elements of information theory}, 
(Wiley, 1991);
John G. Proakis and Masoud Salehi,
{\it Communication systems engineering}, 
(Prentice-Hall, 1994).
%
\bibitem{HC67}J.H. Havrda and F. Charvat,
Kybernetika, vol. 3, pp. 30-35, 1967.
%
\bibitem{Da70}Z. Dar\'oczi,
Inf. Control., vol. 16, pp. 36-51, 1970.
%
\end{thebibliography}
\end{document}